\documentclass[authoryear,preprint,review,12pt]{elsarticle_arxiv}

\usepackage{amssymb,amsmath}
\usepackage{bm}
\usepackage[doublespacing]{setspace}
\usepackage{rotating}
\journal{Icarus}

\begin{document}

\begin{frontmatter}

  \title{High precision comet trajectory estimates: the
    Mars flyby of C/2013~A1 (Siding Spring)}

  \author[jpl]{D. Farnocchia}
  \ead{Davide.Farnocchia@jpl.nasa.gov}
  \author[jpl]{S.~R. Chesley}
  \author[esa,spacedys,iaps]{M. Micheli}
  \author[delamere]{A. Delamere}
  \author[lpl]{R.~S. Heyd}
  \author[ifa]{D.~J. Tholen}
  \author[jpl]{J.~D. Giorgini}
  \author[jpl]{W.~M. Owen}
  \author[jpl]{L.~K. Tamppari}
  
  \address[jpl]{Jet Propulsion Laboratory, California Institute of
    Technology, Pasadena, CA 91109, USA}
  \address[esa]{ESA NEO Coordination Centre, 00044 Frascati (RM), Italy}
  \address[spacedys]{SpaceDyS s.r.l., 56023 Cascina (PI), Italy}
  \address[iaps]{INAF-IAPS, 00133 Roma (RM), Italy}
  \address[delamere]{Delamere Support Services, Boulder, CO 80304,
    USA}
  \address[lpl]{Planetary Image Research Laboratory, Lunar and
    Planetary Laboratory, University of Arizona, Tucson, AZ 85721, USA}
 \address[ifa]{Institute for Astronomy, University of Hawaii, Honolulu, HI 96822, USA}
  
\begin{abstract}
The Mars flyby of C/2013 A1 (Siding Spring) represented a unique opportunity for imaging a long-period comet and resolving its nucleus and rotation period.
Because of the small encounter distance and the high relative velocity, the goal of successfully observing C/2013 A1 from the Mars orbiting spacecrafts posed strict accuracy requirements on the comet's ephemerides.
These requirements were hard to meet, as comets are known for being highly unpredictable: astrometric observations can be significantly biased and nongravitational perturbations affect comet trajectories. Therefore, even prior to the encounter, we remeasured a couple of hundred astrometric images obtained with ground-based and Earth-orbiting telescopes. We also observed the comet with the Mars Reconnaissance Orbiter's High Resolution Imaging Science Experiment (HiRISE) camera on 2014 October 7.
In particular, these HiRISE observations were decisive in securing the trajectory and revealed that out-of-plane nongravitational perturbations were larger than previously assumed.
Though the resulting ephemeris predictions for the Mars encounter allowed observations of the comet from the Mars orbiting spacecrafts, post-encounter observations show a discrepancy with the pre-encounter trajectory.
We reconcile this discrepancy by employing the Rotating Jet Model, which is a higher fidelity model for nongravitational perturbations and provides an estimate of C/2013~A1's spin pole.
\end{abstract}


\end{frontmatter}

\section{Introduction}
On 2014 October 19 long-period comet C/2013~A1 (Siding Spring) experienced an exceptionally close encounter with Mars at $140496.6 \pm 4.0$ km and a relative velocity of $55.963249 \pm 0.000028$ km/s (1$\sigma$ formal uncertainties).
While an impact between the nucleus of C/2013~A1 and Mars had been ruled out by earlier observational data, early predictions by \citet{v_css14} and \citet{m_css14} suggested that the dust in the comet's tail could have posed a significant hazard to the spacecrafts orbiting Mars. 
Later studies \citep{f_css14, k_css14, t_css14, y_css14} used observations of C/2013~A1 to estimate the dust production rate and concluded that spacecrafts were unlikely to be affected by the Mars flyby of C/2013~A1.
Moreover, these studies predicted the fluence peak timing as well as the direction of the incoming dust to inform the rephasing of the spacecraft orbits.
As expected, no damage was reported.

Along with the potential hazard, the close encounter between Mars and C/2013~A1 represented a unique scientific opportunity: never had a long-period comet been observed at such small distance.
The instruments on board of the Mars orbiting spacecrafts could be used to collect valuable observations of the comet and for the first time allow one to resolve the nucleus and rotation period of a long-period comet.

The goal of observing C/2013 A1 during the Mars encounter posed strict requirements on the ephemeris accuracy. While the Mars Reconnaissance Orbiter (MRO) High Resolution Imaging Science Experiment (HiRISE) camera has a very large field of view capability \citep{hirise}, only a small portion of its field of view, 4 mrad $\times$ 4 mrad, could be used for comet observations because of data handling limitations. This translates to a 280 km error tolerance at a distance of 140,000 km. Meeting these requirements was not an easy task as comets are famously difficult to predict.\footnote{\emph{Comets are like cats; they have tails, and they do precisely what they want}, 
David H. Levy}
As a matter of fact, because of comet tails and outgassing, comet astrometric observational data are not as clean as for other bodies, e.g., asteroids.
Moreover, nongravitational perturbations often limit the capability of providing accurate ephemerides \citep{yeomans_comets}: these perturbations can be hard to model and estimate due to the lack of information and yet significantly affect comet trajectories.

Comet ephemeris challenges associated with the Deep Space 1 mission's encounter with Comet 81P/Borrelly in 2001 \citep{rayman} led to the development of a rotating jet model for nongravitational accelerations \citep{comet_ng_chesley}, which provided far superior comet ephemeris predictions, albeit ex post facto. Later, the ephemeris development efforts for subsequent NASA comet flyby missions, such as Deep Impact, Stardust, EPOXI and Stardust-NExT, went smoothly, in large part because of the improvements in nongravitational acceleration modeling \citep[see, e.g.,][]{chesley_mission}.

As we shall see, the technical approaches developed during these five comet flyby missions were invaluable in deriving an accurate ephemeris for the Siding Spring encounter of Mars. A persistent challenge in such cases revolves around understanding how astrometric bias affects the measured nucleus position due to the effects of an asymmetric coma. Complicating the situation is the possible presence of nongravitational accelerations that could be affecting the trajectory in ways that are difficult to distinguish from astrometric bias. Separating these two sources of error, nongravitational accelerations and astrometric bias can be helped by carefully examining fits and predictions associated with subsets of the astrometric data set and by taking special care in the astrometric reduction to measure the peak of the condensation in the target's point spread function. Fortunately, as the time to encounter shortens the problem tends to resolve itself due to the combination of longer data arcs, which tends to improve nongravitational acceleration modeling, and shorter mapping times, which reduce the negative effect of modeling errors.

\section{Analysis prior to the encounter}
\begin{figure}
\centerline{\includegraphics[width=10 cm]{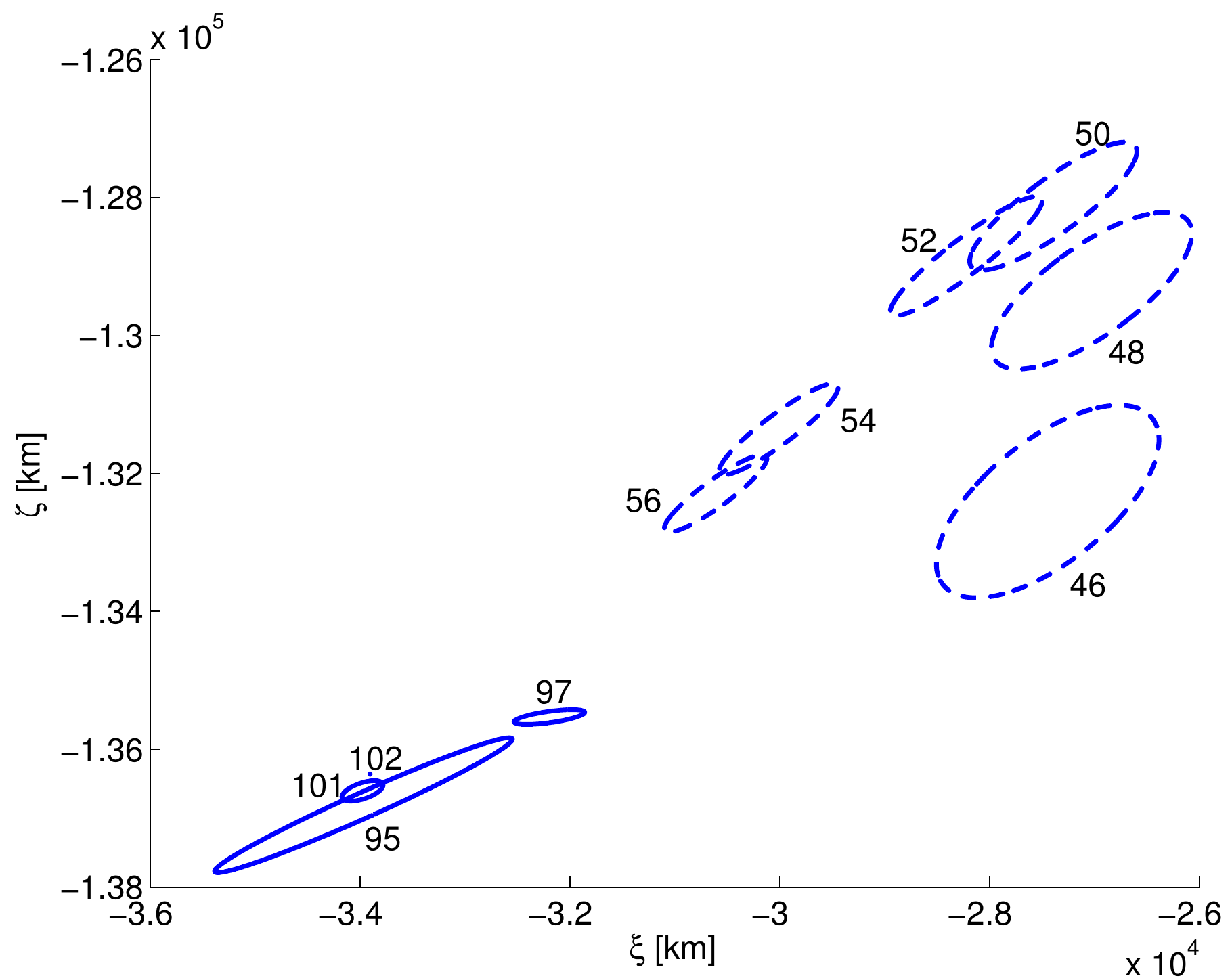}}
\caption{Mapping of the C/2013~A1 1$\sigma$ position uncertainty on the 2014 October 19 $b$-plane for sequential orbital solutions. Dashed ellipses are for earlier gravity-only solutions computed before the astrometry selection and remeasurement process described in Sec.~\ref{s:ast_pre}: solution 46 includes data until 2014 March 31, while solution 56 includes data until 2014 August 11. Solution 95 and 97 are the two solutions computed before the 2014 October 7 HiRISE astrometry, while solution 101 is the final ephemeris delivery prior to the encounter. Finally, solution 102 is computed employing the Rotating Jet Model and using the full data arc through end of April 2015.}\label{f:bplane}
\end{figure}

At the end of March 2014, C/2013~A1 went into solar conjunction thus becoming unobservable from Earth.
The trajectory obtained with the pre-conjunction observed arc was JPL solution 46, which is fully described in \citet{f_css14}.
At the end of May 2014 C/2013~A1 emerged from the Sun, thus leading to additional astrometric observations and corresponding orbit updates.
However, these orbit updates showed a quite erratic behavior.
For different orbital solutions, Fig.~\ref{f:bplane} shows the predicted locations on the 2014 October 19 encounter $b$-plane, which is the plane containing the Mars center at the coordinate origin and normal to the asymptotic incoming velocity of the comet $\mathbf v_\infty$ \citep{bplane}. The projection of the heliocentric velocity of Mars on the $b$-plane is oriented as $-\bm\zeta$ while the $\xi$ coordinate completes the reference frame, i.e., $\bm\xi = \mathbf v_\infty\times \bm\zeta/|\mathbf v_\infty|$ \citep[for more details see][]{valsecchi}. The sequence of orbit updates from solution 46 to solution 56 was statistically inconsistent and raised questions about the ability to provide sufficiently accurate encounter predictions.

\subsection{Astrometry}\label{s:ast_pre}

One of the sources of the erratic behavior shown in Fig.~\ref{f:bplane} is the poor quality of the astrometry used to compute the orbital solutions.
Astrometry of active comets is often affected by an asymmetric distribution of light from the coma surrounding the nucleus.
If a model designed to fit the symmetric light distribution from a point source is used on an active comet, the fitted photocenter can be dependent on the size of the synthetic aperture utilized.

This effect was quite obvious in the astrometry of comet P/Wild 2, obtained in support of the Stardust mission's flyby of that comet \citep{tholen_dps04}. 
As the size of the synthetic aperture increased, the measured photocenter moved in the tailward direction.
At least in the case of P/Wild 2 and over the range of synthetic apertures tested, the offset showed a fairly linear dependence on synthetic aperture size, so the position of the cometary nucleus was computed by linearly extrapolating the fitted coordinates to a theoretical zero-sized synthetic aperture.
The resulting astrometry enabled the uncertainty in the position of P/Wild 2 to be shrunk by a factor of three prior to the spacecraft encounter.

The same tailward bias was observed in the astrometry of C/2013~A1 as the size of the synthetic aperture increased, therefore one should take the astrometric position corresponding to the extrapolation of the trend to a theoretical zero-sized aperture or at least to the brightness peak.
However, we had no control nor information on most of the astrometry that was reported to the Minor Planet Center, some of which obtained by amateur observers.


\begin{sidewaysfigure}
\centerline{\includegraphics[width=20 cm]{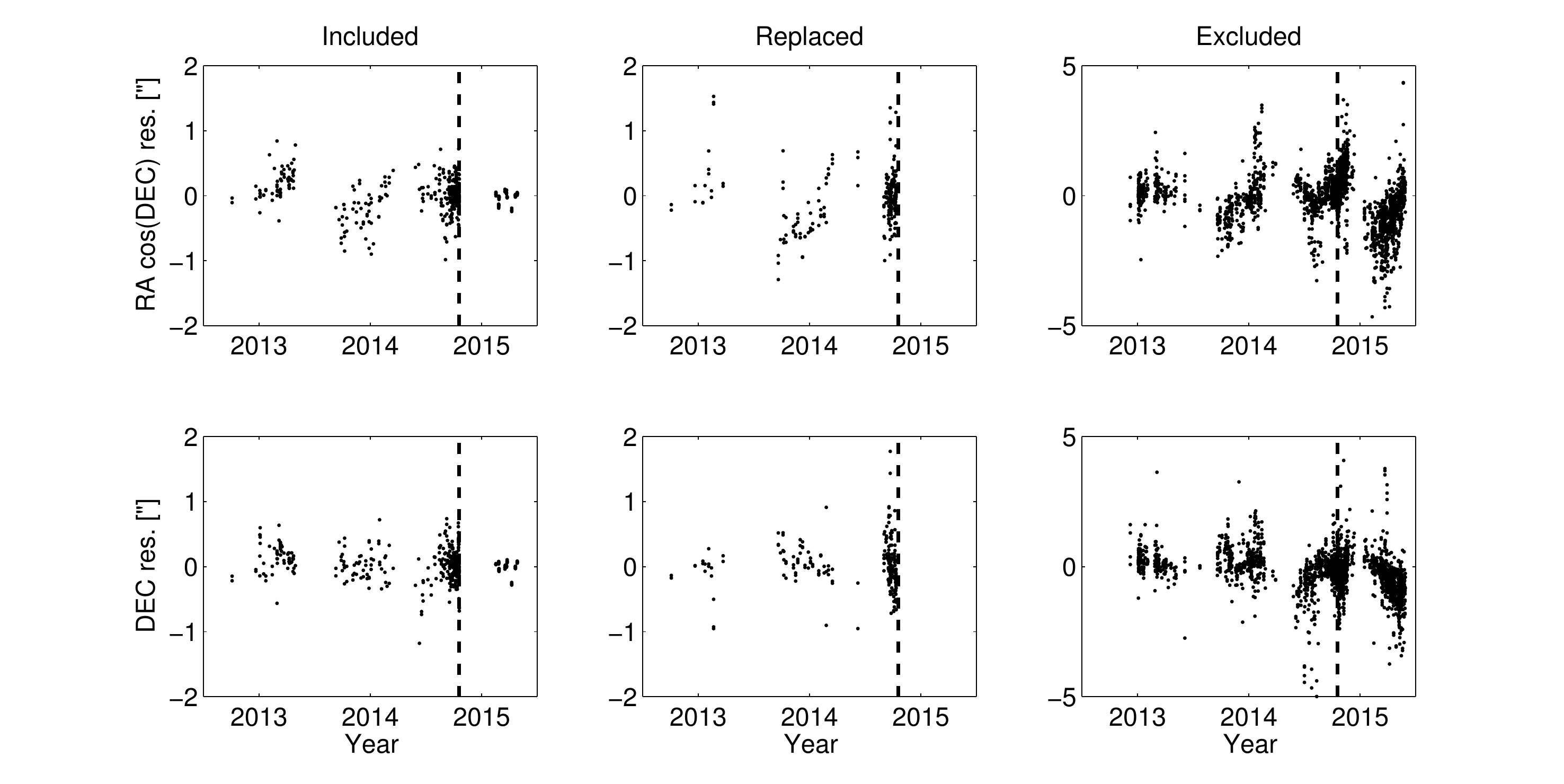}}
\caption{Astrometric residuals of the astrometry until end of April 2015 against solution 102, described in Sec.~\ref{s:s102}. The left column is for the observations included in the fit, the central column for the observations replaced by our remeasurements, and the right column for the observations filtered out by our selection process. The vertical dashed line corresponds to the Mars encounter.}\label{f:res_plot}
\end{sidewaysfigure}

To obtain more reliable orbital solutions, we went through a strict process of astrometry selection and remeasurement.
First, we selected the few astrometric observations obtained by professional surveys: Pan-STARRS 1, Kitt Peak, and Siding Spring Survey.
It is worth pointing out that most of the time C/2013~A1 was at high negative declination, and therefore was not observable from the Northern Hemisphere where most
professional surveys are located.
Then, we added observations that we obtained at the Mauna Kea and Cerro Paranal observatories.

Finally, to obtain a good time coverage and increase the number of observations we asked other observers to provide us with their images so that we could remeasure the astrometric positions prior to the encounter.
We obtained and remeasured more than 200 images obtained at the following observatories:
\begin{itemize}
\item Swift space telescope (this was first time Swift has been used to obtain small body astrometry);
\item RAS Observatory, Moorook;
\item Sutherland-LCOGT A and B;
\item iTelescope Observatory, Siding Spring;
\item Siding Spring-LCOGT A;
\item Blue Mountains Observatory, Leura;
\item CAO, San Pedro de Atacama;
\item Murrumbateman.
\end{itemize}
For these stations, we performed the astrometry with track and stack procedures, and corrected the astrometric positions by as much as 2$''$.
For each image, we estimated the astrometric error from the image quality, following a scale that assigned 1 pixel astrometric error to all the images where an obvious star-like central condensation was visible, and 2 pixels to more problematic cases where the central condensation was either diffuse or blurred (due to poor seeing conditions and/or tracking).
We evaluated a few intermediate cases at 1.5 pixels.
We set the data weights by conservatively applying a scale factor 2 to the estimated uncertainties.
 
Figure~\ref{f:res_plot} shows the right ascension (RA) and declination (DEC)\footnote{In this paper RA and DEC are always referred to the equatorial J2000 reference frame.} residuals against solution 102, which is our current best estimate for the comet's trajectory and is described in Sec.~\ref{s:s102}. 
It is clear how important the remeasurement and observation selection process was: the selected and remeasured astrometric observations have smaller errors than the excluded data.
In particular, some of the observations we filtered out during our selection process have residuals as large as 5$''$, while the astrometric positions replaced by our remeasurements have residuals as large as 2$''$.

\subsection{Dynamical models}
To model nongravitational perturbations we first considered the classical formulation by \citet{comet_ng}:
\begin{equation}\label{e:comet_ng}
\mathbf a_{NG} = g(r) (A_1\hat{\mathbf r} + A_2\hat{\mathbf t} + A_3
\hat{\mathbf n})\ \ ,\ \ g(r) = \alpha \left( \frac{r}{r_0}\right)^{-m} \left[ 1 + \left( \frac{r}{r_0}\right)^{n}\right]^{-k}
\end{equation}
where $r$ is the heliocentric distance, $m = 2.15$, $n = 5.093$, $k =
  4.6142$, $r_0 = 2.808$ au, and $\alpha$ is such that $g(\text{1 au}) =
  1$. $A_1$, $A_2$, and $A_3$ are free parameters that give the
nongravitational acceleration at 1 au in the radial-transverse-normal
reference frame defined by $\hat{\mathbf r} = \mathbf r/r$, $\hat{\mathbf t} = \hat{\mathbf n} \times \hat{\mathbf r}$,
$\hat{\mathbf n} = \mathbf r \times \mathbf v/|\mathbf r \times \mathbf v|$, where $\mathbf r$ and $\mathbf v$ are the heliocentric position and velocity of the comet.
It is common practice to ignore the out-of-plane component, i.e., $A_3 = 0$. 

As of 2014 October 3, it was clear that nongravitational perturbations were needed to fit the observed data.
We computed two different orbital solutions:
\begin{itemize}
\item Solution 95, where $A_1$, $A_2$, and $A_3$ are all determined as part of the least squares orbital fit;
\item Solution 97, where $A_1$ and $A_2$ are determined from the orbital fit and $A_3$ is set to zero.
\end{itemize}
Table~\ref{t:ng_est} shows the values of the nongravitational parameters for solutions 95 and 97, and Fig.~\ref{f:bplane} shows the corresponding $b$-plane predictions.
Solution 95 corresponds to the more conservative approach where no assumptions are made on the nongravitational parameters.
On the other hand, solution 97 is more aggressive, as $A_3$ is set to zero.
The nominal value of $A_3$ for solution 95 is larger than expected, e.g., the reference scenario in \citet{f_css14} corresponded to a $\pm2\times 10^{-9}$ au/d$^2$ 3$\sigma$ interval for $A_3$.
We thought that such a large value of $A_3$ was unlikely and could be due to still unresolved issues in the astrometry.
Moreover, the $A_3$ detection was only marginal (1.3$\sigma$ significance). Therefore, we decided to set $A_3$ to zero and choose solution 97 as baseline ephemeris for the 2014 October 7 HiRISE observations described in Sec.~\ref{s:12dprior}.

\begin{table}
\begin{center}
\begin{tabular}{lcc}
  \hline
Solution & 95 & 97\\
\hline
$A_1$ [$10^{-9}$ au/d$^2$]  & $18.8 \pm 4.8$  & $15.8 \pm 3.5$\\
$A_2$ [$10^{-9}$ au/d$^2$] & $-11.2 \pm 13.0$ & $4.4 \pm 3.5$\\
$A_3$ [$10^{-9}$ au/d$^2$] & $-8.4 \pm 6.5$ & $0 \pm 0$\\
\hline
\end{tabular}
\end{center}
\caption{Estimate of nongravitational parameters (along with 1$\sigma$ formal uncertainties) as of 2014 Oct 3. Solution 95 has all three nongravitational parameters $A_1$, $A_2$, and $A_3$ as free parameters in the fit, while solution 97 is constrained to $A_3 = 0$.}
\label{t:ng_est}
\end{table}

\subsection{HiRISE observations on 2014 October 7}\label{s:12dprior}
The uncertainties in nongravitational perturbations were too large to make sure the comet would be inside the HiRISE field of view for the close approach observations.
For instance, if the pointing was centered on solution 97 then solution 95 would be outside the field of view at closest approach.

To tackle this problem, HiRISE observations from the Mars Reconnaissance Orbiter took place on 2014 October 7.
The much smaller range distance between C/2013~A1 and MRO as well as the different geometry provided much more leverage than Earth-based observations. 
The date for these observations was chosen so that 1) the plane-of-sky uncertainty would be small enough to ensure that C/2013~A1 would be in the field of view; 2) there would be enough time to transmit the images to the ground, perform the astrometric reduction, update the ephemeris and the pointing for the close approach observations; 3) the observations would provide the needed leverage to constrain the trajectory of C/2013~A1.

\begin{figure}
\centerline{\includegraphics[width=12 cm]{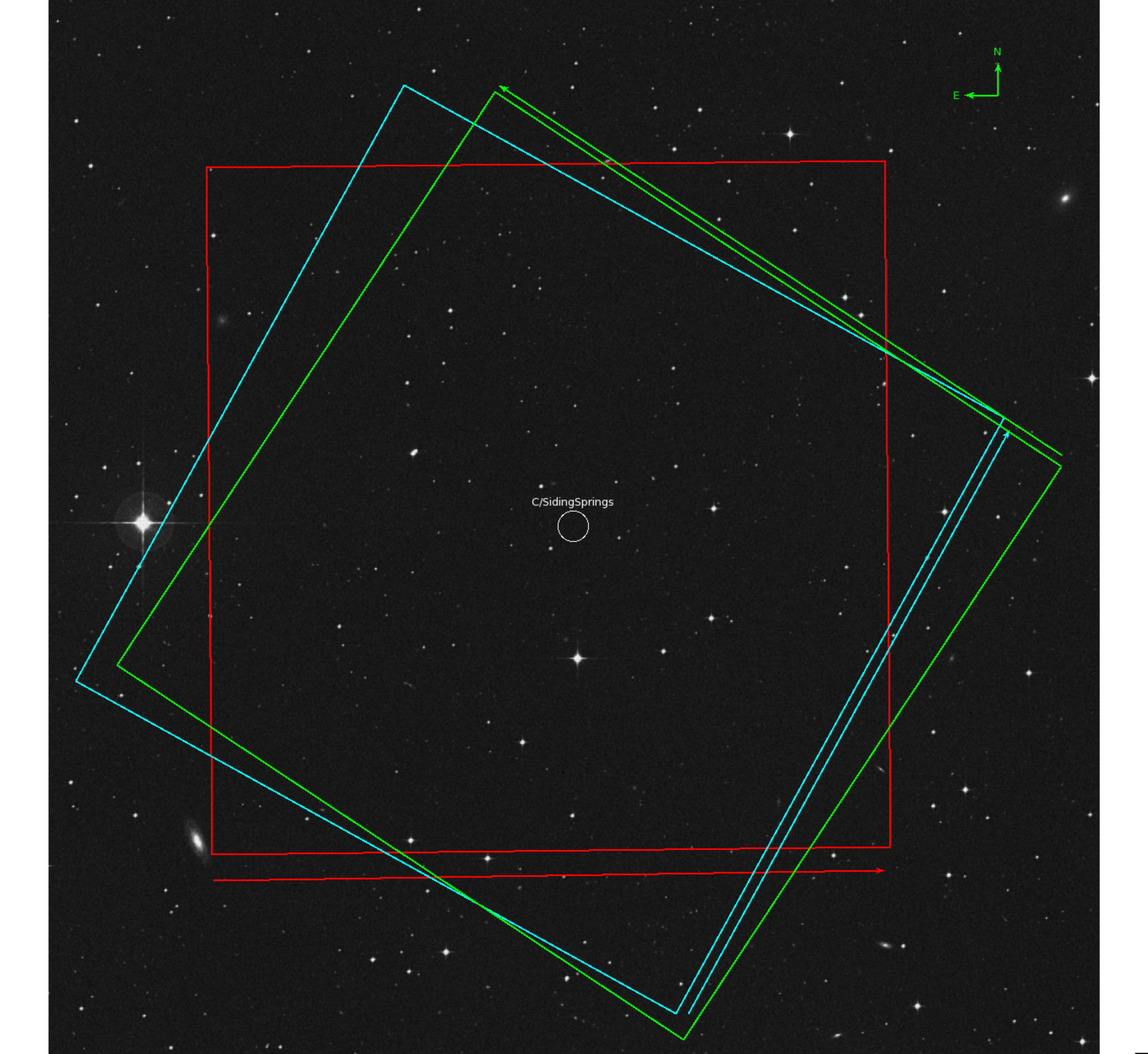}}
\caption{The orientation and scan directions of the three 2014 October 7 observations from HiRISE.  The green, cyan, and red colored boxes are the footprints of the three observations plotted on top of a digital sky survey field. The colored arrows indicate the direction of the scan of the corresponding footprint. The approximate position of C/2013~A1 is indicated in white.}\label{f:hirise_image}
\end{figure}

The HiRISE observations utilized the three central broadband RED CCDs in the instrument focal plane \citep{hirise}, yielding images of the star field surrounding C/2013~A1 that were $\sim$20$'$ on a side as shown in Fig.~\ref{f:hirise_image}.
The scanning rate was selected to obtain the maximum useful integration time of the HiRISE instrument, 2.56 seconds.
The individual CCD images were corrected for dark current, instrument noise and for low frequency spacecraft jitter.
The CCD images were joined  together utilizing reconstructed SPICE kernels provided by NASA's Navigation and Ancillary Information Facility \citep{spice} and SPICE related utilities in the Integrated Software for Imagers and Spectrometers software package \citep{isis}, resulting in a single mosaic of the three CCDs to which an astrometric solution could be applied. To determine the background star positions on the CCD mosaic, their centroids were measured and a plate solution computed from a least squares fit of these star positions.
The position of the comet was determined by measuring the centroid of the comet image and converted into RA and DEC via the plate solution.
Since HiRISE is a scanning instrument, the time of the observation was computed from the image line determined from the centroid measurement.
Figure~\ref{f:hirise_image} shows the orientation and scan direction of the three HiRISE images acquired on 2014 October 7.

The measured astrometric positions are reported in the first block of Table~\ref{t:hirise_ast}.
To account for the uncertainties resulting from centroiding, MRO ephemeris errors, and the small number of reference stars in the field of view, we conservatively weighted these data at 1$''$.
We chose these weights according to an assessment of the internal consistency of the data and the estimated systematic errors.


\begin{table}
\begin{center}
\begin{tabular}{lcccc}
  \hline
Date & RA  & DEC & $\Delta$RA & $\Delta$DEC\\
UTC &  [hh mm ss] & [dd mm ss] & [$''$] & [$''$]\\
\hline
2014-10-07.6273801505 & 02 44 18.0206 & -15 25 39.461 & $+0.212$ & $+0.100$\\
2014-10-07.6447532755 & 02 44 18.1546 & -15 25 27.142 & $+0.006$ & $-0.028$\\
2014-10-07.6621002778 & 02 44 17.2358 & -15 25 16.345 & $-0.630$ & $-0.099$\\
\hline
2014-10-17.2864254282 & 02 44 16.40 & $-$14 54 09.3 & $-0.038$ & $-0.049$\\
2014-10-17.3072505093 & 02 44 15.54 & $-$14 52 34.9 & $+0.084$ & $-0.046$\\
2014-10-17.4421608565 & 02 44 16.16 & $-$14 51 30.4 & $+0.093$ & $+0.670$\\
2014-10-17.4629921412 & 02 44 15.23 & $-$14 49 48.3 & $+0.443$ & $+0.422$\\
2014-10-17.5979258218 & 02 44 15.84 & $-$14 48 29.8 & $-0.372$ & $+0.523$\\
2014-10-17.6187525000 & 02 44 14.83 & $-$14 46 38.4 & $-0.265$ & $+0.614$\\
2014-10-17.7537702431 & 02 44 15.54 & $-$14 45 01.1 & $+0.024$ & $+0.284$\\
2014-10-17.7746015856 & 02 44 14.40 & $-$14 42 58.6 & $-0.113$ & $+0.298$\\
2014-10-17.9094922338 & 02 44 15.16 & $-$14 40 57.8 & $+0.128$ & $-0.049$\\
2014-10-17.9303112616 & 02 44 13.88 & $-$14 38 41.7 & $-0.444$ & $+0.368$\\
2014-10-18.0652698032 & 02 44 14.69 & $-$14 36 09.1 & $-0.241$ & $+0.252$\\
2014-10-18.0860832407 & 02 44 13.32 & $-$14 33 37.2 & $-0.047$ & $+0.394$\\
2014-10-18.2210950694 & 02 44 14.15 & $-$14 30 22.5 & $-0.072$ & $+0.451$\\
2014-10-18.2418975810 & 02 44 12.60 & $-$14 27 30.8 & $-0.246$ & $+0.126$\\
2014-10-18.3768168866 & 02 44 13.49 & $-$14 23 19.4 & $+0.336$ & $+0.075$\\
2014-10-18.3976231829 & 02 44 11.73 & $-$14 20 01.1 & $-0.211$ & $+0.084$\\
2014-10-18.5325476042 & 02 44 12.65 & $-$14 14 28.8 & $+0.190$ & $+0.340$\\
2014-10-18.5533507060 & 02 44 10.71 & $-$14 10 35.9 & $+0.721$ & $+0.435$\\
2014-10-18.6944945486 & 02 44 12.99 & $-$14 01 59.8 & $+0.472$ & $+0.138$\\
2014-10-18.7152678472 & 02 44 05.55 & $-$13 57 05.5 & $-0.456$ & $+0.032$\\
2014-10-18.8507619560 & 02 44 11.78 & $-$13 46 21.1 & $+0.109$ & $+0.122$\\
2014-10-18.8715381481 & 02 44 02.65 & $-$13 40 15.2 & $-0.236$ & $-0.495$\\
2014-10-19.0070050579 & 02 44 10.13 & $-$13 24 17.5 & $+0.262$ & $+0.002$\\
2014-10-19.0277652778 & 02 43 58.53 & $-$13 16 19.1 & $-0.016$ & $+0.029$\\
2014-10-19.1598250347 & 02 44 07.04 & $-$12 52 32.7 & $+0.085$ & $+0.517$\\
2014-10-19.1805583565 & 02 43 56.30 & $-$12 41 15.2 & $+0.059$ & $+0.343$\\
2014-10-19.3161075926 & 02 44 02.73 & $-$11 57 00.0 & $-0.134$ & $-0.253$\\
2014-10-19.3368152315 & 02 43 47.09 & $-$11 38 56.9 & $-0.076$ & $-0.267$\\
2014-10-19.4724035764 & 02 43 54.07 & $-$10 03 09.3 & $+0.114$ & $-0.041$\\
2014-10-20.2152129167 & 14 45 14.70 & $+$19 18 44.9 &$+0.324$ & $-0.639$\\
2014-10-20.3715103356 & 14 45 00.17 & $+$18 20 00.8 &$+0.255$ & $-0.683$\\
\hline
\end{tabular}
\end{center}
\caption{C/2013~A1 astrometry obtained by the HiRISE camera of the Mars Reconnaissance Orbiter. The last two columns contain the observed $-$ computed residuals with respect to solution 102, described in Sec.~\ref{s:s102}. The RA residuals account for the spherical metric factor cos(DEC).}
\label{t:hirise_ast}
\end{table}

\begin{figure}
\centerline{\includegraphics[width=8 cm]{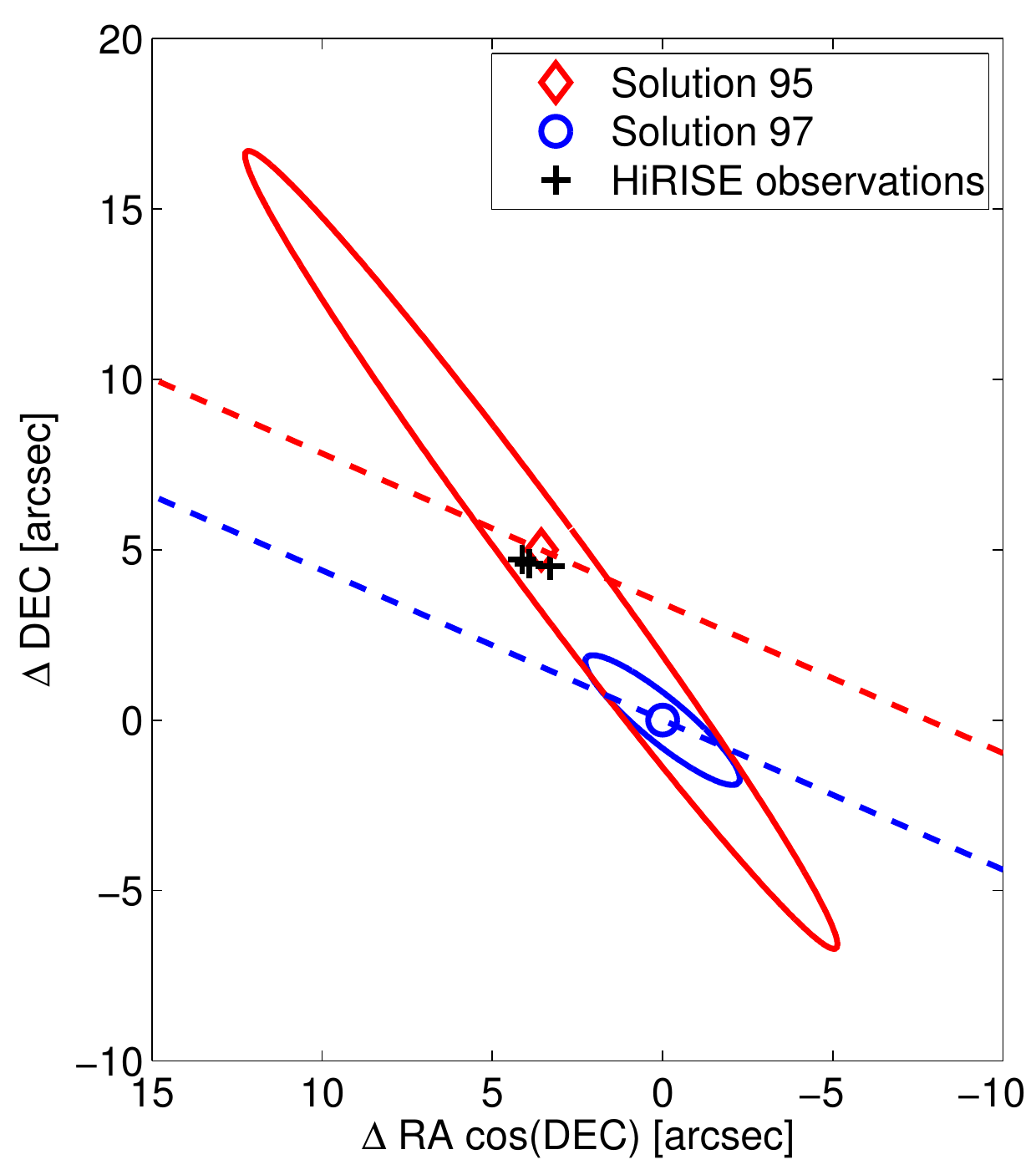}}
\caption{Plane of sky on 2014 October 7 as seen from the Mars Reconnaissance Orbiter. The origin is set on the prediction corresponding to solution 97. Ellipses correspond to the 3$\sigma$ uncertainty level. Dashed curves represent the plane-of-sky motion: C/2013~A1 is moving North-East.}\label{f:mro_97_95}
\end{figure}

Figure~\ref{f:mro_97_95} shows the HiRISE astrometric positions with respect to solutions 95 and 97.
These positions are clearly inconsistent with solution 97, which we had deemed as more likely.
On the other hand, solution 95 provide an excellent prediction for the HiRISE observations, thus confirming the (unexpectedly) large out-of-plane component of the nongravitational perturbations.

The 2014 October 7 HiRISE observations removed the prejudice on $A_3$ by showing that the large out-of-plane perturbation seen in solution 95 was real and not an artifact of unresolved issues in the astrometry.
Since solution 95 and 97 would have led to different field of views for the close approach observing sequence, these HiRISE observations were decisive to obtain accurate ephemerides for the Mars encounter.

\subsection{Ephemerides}
By including the 2014 October 7 HiRISE astrometry as well as some additional ground-based astrometry through 2014 October 13, which was the deadline for the final ephemeris delivery, we computed JPL solution 101.
Table~\ref{t:sol101} shows the corresponding osculating orbital elements, nongravitational parameters, and close approach distance and time. Figure~\ref{f:bplane} shows the corresponding $b$-plane prediction.

Solution 101 was used to successfully observe C/2013~A1 during the close approach to Mars.
The data analysis of the acquired images\footnote{Some of the images are available at http://mars.nasa.gov/comets/sidingspring/images} is still ongoing, though some preliminary results were reported by \citet{delamere_dps}.

\begin{table}
\begin{center}
\begin{tabular}{lc}
 \hline
Epoch TDB & 2014 Feb 3.0\\
Eccentricity & $1.0007424 	 \pm 0.0000029$\\
Perihelion distance & $1.3986359 \pm 0.0000024$ au\\
Time of perihelion TDB & 2014 Oct $25.32823 \pm 0.00019$ d\\
Longitude of node & $300.976279^\circ \pm 0.000048^\circ$\\
Argument of perihelion & $2.44097^\circ \pm 0.00015^\circ$\\
Inclination & $129.027341 \pm 0.000020^\circ$\\
\hline
$A_1$ & $(20.8 \pm 2.6)\times 10^{-9}$ au/d$^2$\\
$A_2$ & $(-13.2 \pm 2.2)\times 10^{-9}$ au/d$^2$\\
$A_3$ & $(-7.4 \pm 1.3)\times 10^{-9}$ au/d$^2$\\
\hline
Mars encounter distance & $140751 \pm 175$ km\\
Epoch of closest approach to Mars TDB & 2014 Oct 19 18:28:38.5 $ \pm\, 1.2$ s\\
\hline
\end{tabular}
\end{center}
\caption{JPL solution 101. The orbital elements are ecliptic heliocentric and error bars correspond to 1$\sigma$ formal uncertainties.}
\label{t:sol101}
\end{table}

\section{Post-encounter reconstruction}

\subsection{Astrometry during and after the encounter}
Besides valuable information on the physical properties of C/2013~A1, the HiRISE observations obtained during the close approach to Mars provide further constraints to the comet's trajectory.
The series of observations taken at closest approach were assembled in a similar fashion to those of the 2014 October 7 HiRISE observations.
Scanning rates and integration times were varied to account for the relative motion of the comet and the spacecraft as well as the increasing brightness of the comet as the distance closed between the spacecraft and the comet.
We could not obtain reliable astrometric measurements for a 20 hour period during closest approach as the scanning rates were too high to allow for accurate measurement of background stars.
The second block of Table~\ref{t:hirise_ast} shows the measured astrometric positions.
Similarly to the 2014 October 7 observations, these data were weighted at 1$''$, with the exception of the last October 18 observation, which was weighted at 2$''$ because of larger RMS errors of the plate solution. 

On 2014 October 19 and 20, C/2013~A1 was observed by Hubble Space Telescope \citep[HST,][]{hst_css}.
We extracted accurate astrometry from drizzle-corrected \citep{drizzle} HST frames
tracked on the comet. Since the reference stars in the field were
trailed, we used trail-fitting techniques to obtain proper
astrometric solutions, referred to the PPMXL catalog \citep{ppmxl}. In each image,
the comet was clearly detected, and the central condensation was sharp
and easy to identify, allowing high-precision astrometric
measurements. To minimize the error from coma asymmetries, we used a small
2-pixel astrometric aperture in all cases.

Because of the high quality of the HST images, the uncertainties of the astrometric positions were within 0.05$''$.
However, when fitting these data we found residuals as large as 0.25$''$.
The cause of these larger residuals is the limited accuracy of the HST ephemeris.
The geocentric coordinates of HST that we used are obtained from JPL's Horizons system \citep{horizons} and are derived from the Two-Line Element (TLE)\footnote{http://spaceflight.nasa.gov/realdata/sightings/SSapplications/Post/JavaSSOP/SSOP\_Help/tle\_def.html} orbit solutions provided by the Joint Space Operation Center\footnote{https://www.space-track.org/auth/login}.
TLEs are not Keplerian elements and are fits to a particular dynamical model called SGP4 \citep{sgp4}.
We analyzed the day-to-day variations by comparing the ephemerides of the HST TLE releases between 2014 October 2 and 17.
The difference between the analyzed solutions suggest that the accuracy of the HST ephemeris is no better than tens of km. 
Therefore, we weighted the HST observations at 0.3$''$.

To extend the observed arc and include astrometry after the Mars flyby, we observed C/2013~A1 from the Mauna Kea observatory from February to April 2015.
While the astrometric positions we reported to the Minor Planet Center are from a 2.0 pixel (0.88$''$) radius aperture, here we adopted the same linear extrapolation of the trend to a theoretical zero-sized aperture to compute the position of the cometary nucleus as desciribed in Sec.~\ref{s:ast_pre}.
Photocenters were computed using synthetic aperture radii ranging from 2.0 to 6.0 pixels in steps of 0.1 pixels, so the extrapolation from 2.0 to 0.0 pixels represents 50 percent of the range actually fit.
An average value for the tailward bias was computed from the trend seen in multiple exposures on each night (see Table~\ref{t:bias_568}).
By default, the symmetric fitting function computes the background level from an annulus surrounding the synthetic aperture with inner and outer radii of 2.5 and 5.0 times the radius of the synthetic aperture, respectively.
For a pixel size of 0.439 arcsec, the background annulus was superposed on the coma of the comet, so some of the light from the comet was being subtracted from the area inside of the synthetic aperture.
As such the reported photometry underestimates the true brightness of the comet for that aperture size.
However, the goal here was accurate astrometry, not accurate photometry.
We conservatively weighted these observations at $0.3''$ to $0.5''$, depending on the size of the bias corrections and the number of observations in the same observing night \citep{fcct15}.

\begin{table}
\begin{center}
\begin{tabular}{lcc}
  \hline
Date & East bias & North bias\\
\hline
2015-02-17 & 0.175$''$ & 0.025$''$\\
2015-02-26 & 0.175$''$ & 0.00$''$\\
2015-02-27 & 0.16$''$ & 0.00$''$\\
2014-03-19 & 0.145$''$ & 0.04$''$\\
2015-03-20 & 0.155$''$ & 0.035$''$\\
2015-03-24 & 0.13$''$ & 0.045$''$\\
2015-03-26 & 0.145$''$ & 0.055$''$\\
2015-04-22 & 0.10$''$ & 0.12$''$\\
2015-04-28 & 0.06$''$ & 0.125$''$\\
\hline
\end{tabular}
\end{center}
\caption{Astrometric corrections for the Mauna Kea post-encounter astrometry. These corrections are obtained by extrapolating to the limit of zero-sized aperture and were added to the positions reported to the MPC.}
\label{t:bias_568}
\end{table}

\subsection{Discrepancy between pre-encounter orbit and post-encounter data}\label{s:disc}
Though solution 101 provided successful ephemeris predictions for the HiRISE close approach observations of C/2013~A1, our attempt to extend the data arc revealed a clear discrepancy between the post-encounter data and the orbital predictions.
For instance, when fitting $A_1$, $A_2$, and $A_3$ to the full data arc, we find $A_3 = (-0.77 \pm 0.15) \times 10^{-9}$ au/d$^2$, which is not statistically compatible with the much stronger negative acceleration measured before the encounter (see Table~\ref{t:sol101}).

To quantify the discrepancy between the pre-encounter solution and the post-encounter data, we measured how much extending the data arc after the encounter changes the orbital solution.
As shown by the fourth row in Table~\ref{t:ng_models}, there is a statistically unacceptable $>$10$\sigma$ correction from the pre-encounter best-fit solution to the full arc best-fit solution. 

\begin{sidewaystable*}
\begin{center}
\begin{tabular}{lc|cccc|cc|cc}
  \hline
Model & N & $\chi^2$ & $\langle{\Delta \text{RA}}\rangle$ & $\langle{\Delta \text{DEC}}\rangle$ & $\sigma$ & $\xi$ & $\zeta$ & Pole & $\Delta$ pole\\
            &   par.          &               &         [$''$]          &  [$''$]                         &     shift   & [km] & [km]         &   (RA, DEC) & \\
\hline
Gravity only & 6 & 16523 & -0.75 & 0.17 & 84.2 & -33853 & -136340 & NA & NA\\
$A_1$ & 7 & 510 & 0.01 & 0.18 & 17.0 & -33900 & -136356 & NA & NA\\
$A_1$, $A_2$ & 8 & 216 & -0.04 & 0.04 & 8.9 & -33900 & -136349 & NA & NA\\
$A_1$, $A_2$, $A_3$ & 9 & 191 & -0.02 & -0.02 & 10.1 & -33899 & -136350 & $(311^\circ, -23^\circ)$ & $115^\circ\ (65^\circ)$\\
$A_1$, $A_2$, $A_3$, $\Delta T$ & 10 & 190 & -0.02 & -0.02 & 10.0 & -33900 & -136350 &$(311^\circ, -23^\circ)$ & $115^\circ\ (65^\circ)$\\
\hline
1 jet & 10 & 274 & 0.06 & 0.12 & 12.7 & -33899 & -136359 & $(141^\circ, 62^\circ)$ & $72^\circ\ (108^\circ)$\\
1 jet $+\ A_1$ & 11 & 89 & 0.01 & -0.02 & 3.2 & -33902 & -136360 & $(80^\circ, 0^\circ)$ & $22^\circ\ (158^\circ)$\\
1 jet $+\ \theta$ & 11 & 165 & 0.01 & 0.04 & 8.5 & -33899 & -136354 & $(350^\circ, 40^\circ)$ & $68^\circ\ (112^\circ)$\\
1 jet $+\ A_1 +\ \theta$ & 12 & 89 & 0.01 & -0.02 & 3.4 & -33903 & -136360 & $(77^\circ, 1^\circ)$ & $19^\circ\ (161^\circ)$\\
2 jets & 12 & 86 & 0.02 & 0.02 & 2.6 & -33902 & -136362 & $(280^\circ, -23^\circ)$ & $144^\circ\ (36^\circ)$\\
2 jets $+\ A_1$ & 13 & 86 & 0.02 & 0.02 & 2.6 & -33902 & -136362 & $(280^\circ, -22^\circ)$ & $144^\circ\ (36^\circ)$\\
2 jets $+\ \theta$ & 13 & 73 & 0.00 & 0.03 & 0.9 & -33904 & -136359 & $(63^\circ, 14^\circ)$ & NA\\
\hline
\end{tabular}
\end{center}
\caption{Orbital fit statistics, $b$-plane coordinates, and spin pole for different nongravitational perturbation models.
The whole arc contains 370 optical and 67 satellite astrometric observations through the end of April 2015. 
$\langle{\Delta \text{RA}}\rangle$ and $\langle{\Delta \text{DEC}}\rangle$ are the average residuals in $\text{RA}\cos(\text{DEC})$ and DEC.
The sixth column contains the $\sigma$ shift between the pre-encounter solution and the whole arc solution, i.e., $\sigma^2 = (x_\text{full} - x_\text{pre})^T \Gamma_\text{pre}^{-1}(x_\text{full} - x_\text{pre})$ where $x_\text{pre}$ is the pre-encounter best fitting solution, $\Gamma_\text{pre}$ its covariance matrix, and $x_\text{full}$ the full arc best fitting solution. The last column shows the angular distance between the considered model's pole and that of the two jets $+\ \theta$ model (numbers in parentheses are for the diametrical opposite pole).}
\label{t:ng_models}
\end{sidewaystable*}


\subsection{Dynamical models}
The observed discrepancy points to the need of refining the nongravitational perturbation model.
The \citet{comet_ng} model does a good job at fitting the pre-encounter observations of C/2013~A1, actually none of the other models discussed below improves the fit to the pre-encounter arc. However, this model  proves inadequate as the trajectory is further constrained by post-encounter data.

We now analyze different nongravitational perturbation models, whose results are summarized in Table~\ref{t:ng_models}.
Besides the orbital fit statistics, to quantify the predictive power of the models Table~\ref{t:ng_models} shows the $\sigma$ shift applied to the pre-encounter solution to fit the post-encounter data.

As described by \citet{dt_ng}, the first refinement is to introduce an offset $\Delta T$ of the activity relative to perihelion, which is where the \citet{comet_ng} nongravitational perturbations peak. This asymmetry can be obtained by replacing $g(r(t))$ with $g(r(t + \Delta T))$ in Eq.~\eqref{e:comet_ng}. For C/2013~A1 the inclusion of $\Delta T$ does not result in any significant improvement: $\chi^2$ is essentially unchanged, the $\Delta T$ estimate from the orbital fit is $1.6 \pm 2.8$ d, which is consistent with $\Delta T = 0$, and we still have a 10$\sigma$ shift from pre-encounter best-fit solution to post-encounter best-fit solution.

A higher fidelity model is the Rotating Jet Model (RJM), which is fully described in \citet{comet_ng_chesley} and we briefly recall here. 
The RJM considers the action of discrete jets on a rotating nucleus.
Given a jet, we compute the mean acceleration averaged over a comet rotation.
The acceleration depends on the orientation of the spin pole, e.g., defined by RA and DEC, the jet's thrust angle $\eta$, i.e., the colatitude of the location of the jet, and the strength $A_J$ of the jet.
The model can be extended to include a diurnal lag angle $\theta$ and thus account for diurnal lag in the jet's activity.
All these parameters can be estimated as part of the orbital fit to the available astrometry.
Note that the RJM does not depend on the rotation period and, unless $\theta \neq 0^\circ$, is insensitive to the sense of rotation.

It is worth pointing out that \citet{li} detected two jet-like features in HST images obtained on 2013 October 29, 2014 January 21, and 2014 March 11, which therefore support the application of the RJM.

The RJM with one jet does not work well, e.g., $\chi^2$ is worse than that obtained with the \citet{comet_ng} model.
This behavior is somehow expected: as the comet moves around the Sun the latitude of the sub-Solar point changes and at some point the jet gets confined in the comet's shadow for the full rotation period, thus turning off the corresponding nongravitational accelerations.

A possible refinement to the single jet model is obtained by adding $A_1$, so that $A_1$ gives the main nongravitational perturbation in the radial direction while the jet mostly contributes to the other components.
This configuration results in a significantly lower  $\chi^2$ value and the $\sigma$ shift is 3.2, which is perfectly reasonable in an 11-dimensional parameter space (there is a 51\% probability that $\Delta \chi^2 > 3.2^2$ with eleven degrees of freedom).

The improvement obtained by adding $A_1$ warrants the addition of a second jet to the model.
After all, using two jets is a sort of generalization of the one jet $+\ A_1$ solution.
The orbital fit statistics and the $\sigma$ shifts of these two models are essentially equivalent.

\begin{figure}
\centerline{\includegraphics[width=8 cm]{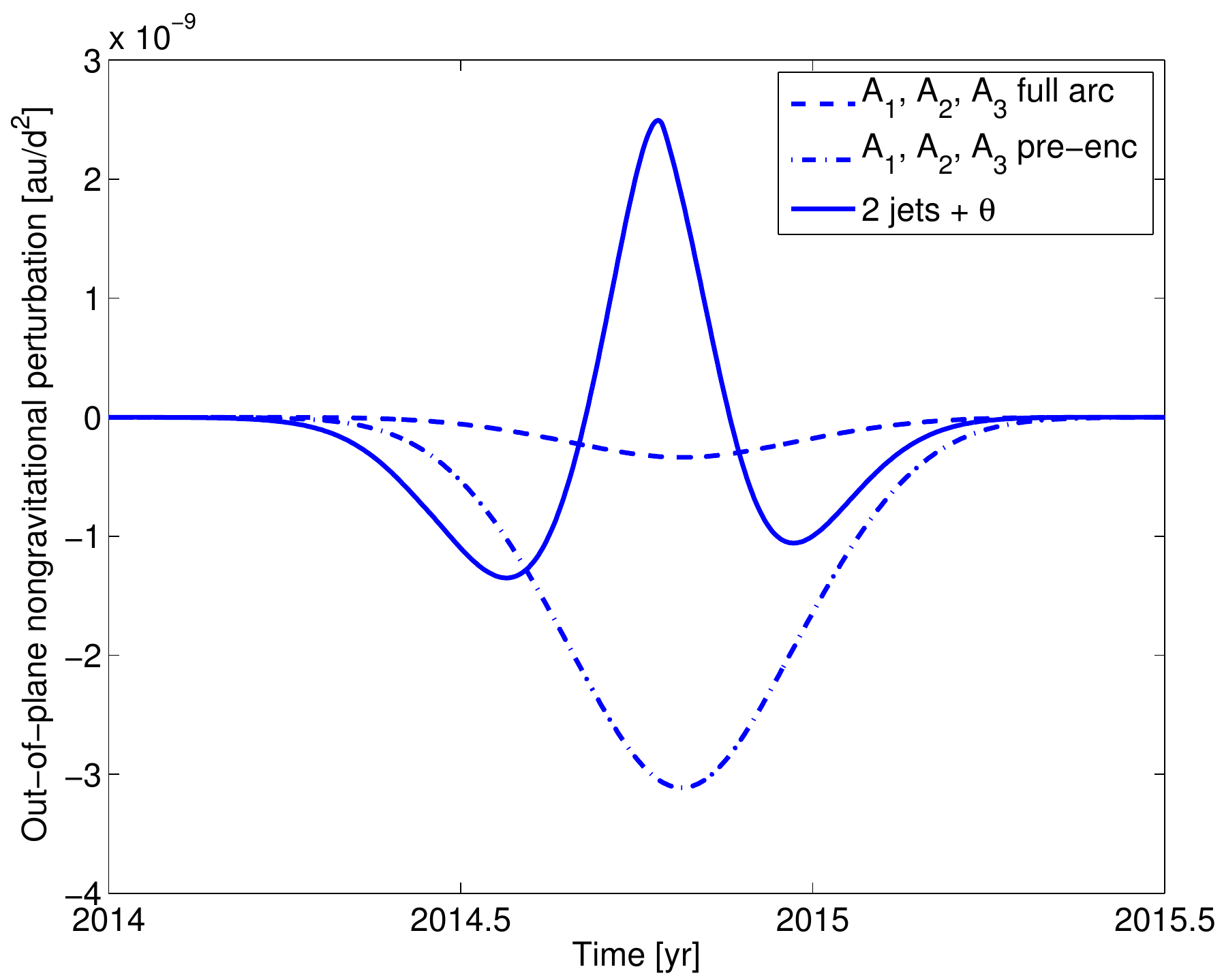}}
\caption{Out-of-plane nongravitational perturbation for three orbital solutions: 1) \citet{comet_ng} model, best fit to the full arc,  2) \citet{comet_ng} model, best fit to the pre-encounter arc, and 3) RJM model with two jets and diurnal lag angle, best fit to the full arc.}\label{f:normal_ng}
\end{figure}

Further adding $A_1$ to the two jets solution does not provide any improvement, thus suggesting that a third jet is not needed.
On the other hand, the inclusion of the diurnal lag angle $\theta$ lowers $\chi^2$ by 13 with respect to the best-fit two jets solution, which is a significant change for the addition of a single parameter.
Moreover, the two jets $+\ \theta$ solution has the lowest $\sigma$ shift when the arc is extended after the encounter, which means that pre-encounter predictions are consistent with post-encounter data.
This consistency can also be seen in the out-of-plane nongravitational perturbations, as shown in Fig.~\ref{f:normal_ng}.
The large difference between the pre-encounter and the full arc best-fit solutions obtained with the \citet{comet_ng} model is evident: the pre-encounter solution has much stronger out-of-plane perturbations, as already discussed in Sec.~\ref{s:disc}.
The RJM with two jets $+\ \theta$ allows short-term variations due to the rapidly changing orientation of C/2013~A1 with respect to the Sun. In particular, around the encounter one of the jets ends up in the comet's shadow and becomes inactive.
Thanks to the change of sign of the RJM out-of-plane perturbation, we can match both the large out-of-plane perturbations prior to the encounter as well as the much lower (in absolute value) average value over the full observed arc.
 
In conclusion, the fit to the observational data and the HST images in \citet{li} warrant the usage of the RJM with two jets.
Moreover, the fit also points to a diurnal lag in the jets' activity.
Interestingly, there is observational evidence supporting the presence of such a lag for comets: measurements by the MIRO instrument on Rosetta show that for 67P/Churyumov-Gerasimenko there is a displacement between the sub-surface temperature peak and the sub-solar point \citep{gulkis}. 
 
\subsection{Ephemerides}\label{s:s102}
As our best estimate of C/2013~A1's trajectory we computed JPL solution 102 by using the two jets $+\ \theta$ model for nongravitational perturbations. 
Table~\ref{t:sol102} shows the corresponding orbital elements, RJM parameters, and close approach data along with their formal uncertainties.

\begin{table}
\begin{center}
\begin{tabular}{lc}
 \hline
Epoch TDB & 2014 Jul 18.0\\
Eccentricity & $1.0008169 \pm 0.0000016$\\
Perihelion distance & $1.39866662 \pm 0.00000071$ au\\
Time of perihelion TDB & 2014 Oct $25.318385 \pm 0.000096$ d\\
Longitude of node & $300.977577^\circ \pm 0.000026^\circ$\\
Argument of perihelion & $2.438202^\circ \pm 0.000060^\circ$\\
Inclination & $129.027337 \pm 0.000031^\circ$\\
\hline
RA of spin pole & $62.8^\circ \pm 7.0^\circ$\\
DEC of spin pole & $14^\circ \pm 14^\circ$\\
Strength of jet 1 & $(86.5 \pm 5.2) \times 10^{-9}$ au/d$^2$\\
Strength of jet 2 & $(153 \pm 67) \times 10^{-9}$ au/d$^2$\\
Thrust angle of jet 1 & $60.2^\circ \pm 3.1^\circ$\\
Thrust angle of jet 2 & $152.4^\circ \pm 4.8^\circ$\\
Diurnal lag angle & $9.4^\circ \pm 1.6^\circ$\\
\hline
Mars encounter distance & $ 140496.6 \pm 4.0$ km\\
Epoch of closest approach to Mars TDB & 2014 Oct 19 18:28:33.99 $ \pm\, 0.85$ s\\
\hline
\end{tabular}
\end{center}
\caption{JPL solution 102. The orbital elements are ecliptic heliocentric and error bars correspond to 1$\sigma$ formal uncertainties.}
\label{t:sol102}
\end{table}

Solution 102 provides a good fit to the astrometric data, e.g., the square root of the reduced $\chi^2$ is 0.29. 
In particular, Table~\ref{t:hirise_ast} contains the HiRISE astrometry residuals with respect to solution 102, which are well within the uncertainties used to weight the data, 2$''$ for last October 18 observation, 1$''$ otherwise. The left column of Fig.~\ref{f:res_plot} shows the RA and DEC residuals again solution 102 as a function of the observation epoch.


Table~\ref{t:ng_models} reports the $b$-plane coordinates $\xi$ and $\zeta$ for solution 102, which are also shown in Fig.~\ref{f:bplane}, as well as those obtained with other dynamical models.
The formal uncertainties in $\xi$ and $\zeta$ are $\sim$4 km, while using the \citet{comet_ng} nongravitational perturbation model can cause differences in $\zeta$ as large as $\sim$10 km.
Despite the small encounter distance, 140500 km, the Mars flyby caused modest changes to the orbit of C/2013~A1 because of the high encounter velocity of 56 km/s and the relatively small mass of the planet.
For instance, the angle between the incoming and outgoing relative velocities of the comet was only $0.01^\circ$.

Interestingly, solution 102 comes with an estimate of C/2013~A1's spin pole.
While this pole allows us to fit the data and accurately describe the comet's trajectory, it is important to point out that this estimate is model dependent and strongly relies on the accuracy of the RJM and its assumptions, e.g., number of jets, simple rotation, etc.
Table~\ref{t:ng_models} shows the different estimates of the spin pole with the \citet{comet_ng} model \citep[spin pole approximated by the direction of the $(A_1, A_2, A_3)$ vector at time $\Delta T$, as done in][]{comet_ng_chesley} and the different configurations of the RJM.
In particular, setting the diurnal lag angle to $0^\circ$ moves the pole by $36^\circ$.

\begin{figure}
\centerline{\includegraphics[width=10 cm]{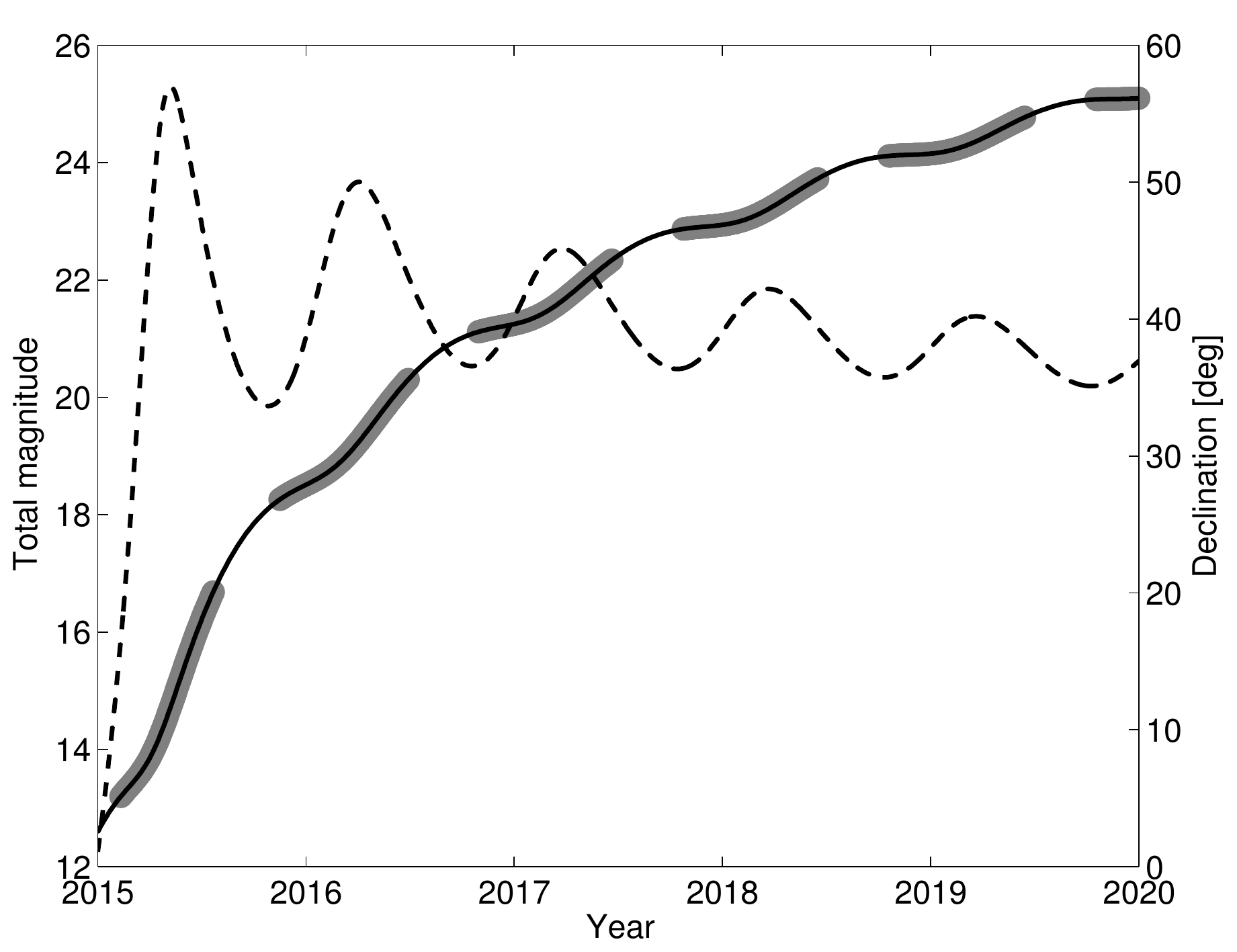}}
\caption{Predicted magnitude (solid line) and declination (dashed line) as observed from the geocenter. Grayed parts correspond to solar elongation $< 60^\circ$.}\label{f:mag_pred}
\end{figure}

Comet C/2013~A1 is still observable in the northern sky.
Figure~\ref{f:mag_pred} shows the predicted total magnitude and declination until 2020, when C/2013~A1 will be at more than 15 au from the Sun.
The next few years are the last chance of observing the comet for quite some time. In fact, the post-perihelion orbit of C/2013~A1 with respect to the Solar System barycenter is elliptic and has an orbital period of more than 500000 years.

\section{Conclusions}
The campaign for observing C/2013~A1 from the Mars orbiting spacecrafts during the 2014 October 19 encounter was challenging in terms of computing reliable comet trajectories.
The small field of view of Mars Reconnaissance Orbiter's HiRISE camera and the close encounter distance set tight requirements on the ephemeris accuracy.

One of the first problems we had to face was the generally poor quality of C/2013~A1's astrometric observations.
In fact, comet astrometry is often affected by tailward biases as the center of light does not generally correspond to the brightness peak.
To mitigate this issue, we first applied a strict filter to the astrometric data by only selecting observations coming from professional surveys.
Then, we added observations for which we directly performed the astrometric reduction either because we directly performed the observations or the original observers provided us with their astrometric images.
This process allowed a more stable orbit update sequence as new data were available.

Nongravitational accelerations often are the limiting factor in terms of comet ephemeris accuracy.
To constrain the nongravitational perturbations acting on C/2013~A1 we planned observations from the Mars Reconnaissance Orbiter HiRISE camera twelve days before the Mars encounter.
Thanks to the favorable geometry, these observations were decisive to secure the trajectory of C/2013~A1 and to provide accurate ephemerides for the encounter.
In particular, the HiRISE observations clearly revealed a significant out-of-plane nongravitational acceleration thus showing how the common practice of restricting to in-plane perturbations can lead to inaccurate predictions.

Though we only have astrometric data for a single revolution, the good observational coverage and the strong leverage provided by the HiRISE observations acquired as the comet flew by Mars showed that the classical \citet{comet_ng} model for nongravitational perturbations is not accurate enough to model the trajectory of C/2013~A1 over the full observed arc.
We therefore resorted to the Rotating Jet Model (RJM), which is a higher fidelity model that allows us to nicely fit the full observed arc thanks to the action of two jets.
Moreover, the RJM provides an estimate of the spin pole of C/2013~A1. 
Though the reliability of this estimate strongly depends on the accuracy of the model, \citet{comet_ng_chesley} already used the RJM to quite successfully estimate the spin pole of 67P/Churyumov-Gerasimenko: the difference between prediction and the actual spin pole orientation \citep{rosetta} is only $13^\circ$.
The analysis of the images acquired during the close encounter will possibly provide an independent estimate of the spin pole to be compared to that obtained from the orbital dynamics.

Future work will be related to further refinements of the nongravitational perturbation.
For instance, the standard $g(r)$ function used by the \citet{comet_ng} model and the RJM is driven by the sublimation rate of water ice as a function of heliocentric distance.
Directly estimating the H$_2$O and CO$_2$ production rates of C/2013~A1 and future astrometric observations could allow one to derive a more C/2013~A1 specific $g(r)$.

\section*{Acknowledgments}
The authors are grateful D. Bodewits, L. Buzzi, P. Camilleri,
D. Herald, K.~R. Hills, E.~I. Kardasis, J.-L. Li, A. Maury,
M.~J. Mattiazzo, J. Oey, J.-F. Soulier, D. Storey, H.~B. Throop, and
H. Williams for sharing their astrometric images. We thank
M.~S.~P. Kelley and C.~M. Lisse for supporting our request of astrometric
images.


Part of this research was conducted at the Jet Propulsion Laboratory,
California Institute of Technology, under a contract with NASA.

Copyright 2015 California Institute of Technology.

\end{document}